\documentclass[%
superscriptaddress,
 amsmath,amssymb,
 aps,
 reprint,
floatfix,
longbibliography
]{revtex4-1}

\usepackage{graphicx}
\usepackage{dcolumn}
\usepackage{bm}
\usepackage[T1]{fontenc}

\AtBeginDocument{%
    \newwrite\bibnotes
    \def\bibnotesext{Notes.bib}
    \immediate\openout\bibnotes=\jobname\bibnotesext
    \immediate\write\bibnotes{@CONTROL{REVTEX41Control}}
    \immediate\write\bibnotes{@CONTROL{%
    apsrev41Control,author="08",editor="1",pages="1",title="0",year="1"}}
     \if@filesw
     \immediate\write\@auxout{\string\citation{apsrev41Control}}%
    \fi
}%

\begin{document}


\title{Striation lines in intermittent fatigue crack growth in an Al alloy}

\author{Anniina Kinnunen}
\affiliation{Department of Applied Physics, Aalto University, P.O. Box 11100, 00076 Aalto, Espoo, Finland}
\author{Ivan V. Lomakin}
\email{Corresponding author: ivan.lomakin@aalto.fi}
\affiliation{Department of Applied Physics, Aalto University, P.O. Box 11100, 00076 Aalto, Espoo, Finland}
\author{Tero M\"{a}kinen}%
\affiliation{Department of Applied Physics, Aalto University, P.O. Box 11100, 00076 Aalto, Espoo, Finland}
\affiliation{NOMATEN Centre of Excellence, National Centre for Nuclear Research, A. Soltana 7, 05-400  Otwock-Świerk, Poland}
\author{Kim Widell}
\affiliation{Department of Mechanical Engineering, Aalto University, P.O. Box 11000, 00076 Aalto, Espoo, Finland}
\author{Juha Koivisto}
\affiliation{Department of Applied Physics, Aalto University, P.O. Box 11100, 00076 Aalto, Espoo, Finland}
\author{Mikko J. Alava}
\affiliation{Department of Applied Physics, Aalto University, P.O. Box 11100, 00076 Aalto, Espoo, Finland}
\affiliation{NOMATEN Centre of Excellence, National Centre for Nuclear Research, A. Soltana 7, 05-400  Otwock-Świerk, Poland}

\date{\today}

\begin{abstract}
Fatigue failure of crystalline materials is a difficult problem in science and engineering, and recent results have shown that fatigue crack growth can occur in intermittent jumps which have fat-tailed distributions.
As fatigue crack propagation is known to leave markings -- called striations -- on the fracture surface, the distances between these should also have fat-tailed distributions, if the crack propagation is intermittent. 
Here, we combine macroscale crack tip tracking in fatigue crack growth measurements of aluminum 5005 samples with \emph{post-mortem} scanning electron microscopy imaging of the striation lines. We introduce two different methods for extracting the striation line spacing from the images.
What we find is a similar distribution of striation spacings as jump sizes using one of our methods, but the average striation spacing does not correlate with the crack growth rate.
We conclude that we observe avalanche-like crack propagation, reflected in both the macroscale crack tip tracking as well as the analysis of the fracture surfaces.
Our results show that the fracture surfaces can be used to study the intermittency of fatigue crack propagation and in development of crack-resistant materials. The advantages and disadvantages of the two methods introduced are discussed.
\end{abstract}

\maketitle

\section{Introduction}

Fatigue is an important physics problem on multiple scales, ranging from the atomistic to the failure of engineering structures. 
Fatigue failure of ductile materials~\cite{ritchie1999mechanisms}, such as metals, can in laboratory conditions be studied by tracking the macroscopic crack growth using various methods, but in engineering applications this is rarely a possibility. Instead one might have to rely on the analysis of \emph{post-mortem} fracture surfaces~\cite{ponson2007crack} to gain useful insights into the fatigue failure processes.
However the stochastic nature of some microscopic observables raises questions about the predictability of crack growth.
An obvious question is then: can one connect the statistics of microscopic observables to the behavior of the macroscopic crack growth rate?\\

The macroscopic crack growth rate can be observed with various methods and one can then fit crack growth laws, such as the Paris--Erdogan law~\cite{paris1963critical}
\begin{equation} \label{eq:paris}
    \frac{\mathrm{d} a}{\mathrm{d} N} = C \Delta K^m
\end{equation}
where $a$ is the crack length, $N$ the number of loading cycles, $C$ a material- (and loading-) specific constant, $K$ the stress intensity factor (SIF) range, and $m$ the material- (and loading-) specific crack growth exponent. The SIF range can be written in terms of the peak SIF $\Delta K = (1-R) K_{max}$ using the stress ratio $R = K_{min}/K_{max}$.
The power-law form of Eq.~\ref{eq:paris} points to the direction of apparent self-similarity~\cite{ritchie2005incomplete, carpinteri2007self} of fatigue crack growth.\\

The determination of the crack length and therefore the crack growth rate is in most cases difficult, and the methods used are either prone to errors or reduced to consider only one corner of the crack. 
Fractography provides an alternative look into the same problem, from the viewpoint that the topography of a \emph{post-mortem} fracture surface is a characteristic of the sample microstructure and the test conditions~\cite{hull1999fractography}, such as the different stages of crack propagation~\cite{oborin2016multiscale,bannikov2016experimental,naimark2021critical,dharmadhikari2021dual}. Extraction of features and characteristics of the fracture surface, so called quantitative fractography~\cite{Underwood1986}, enables the study of the sizes, shapes, orientations and other measurable values related to the features.

For fatigue the most important surface features are the periodic traces -- first time reported as a “platy patterns” on the fracture surface~\cite{zapffe1951platypatterns} observed at a high magnification using scanning electron microscopy~(SEM). These traces we originally called slip bands by Thompson and Wadsworth~\cite{thompson1958metal} and later striations by Nine and Kuhlmann-Wilsdorf~\cite{nine1957cjp}. The striations are lines generated by the advancement of the fracture line and they are generally suggested as an experimental technique to probe the properties of the fatigue crack propagation.\\

The general view has been that fatigue crack growth (especially in the Paris regime where Eq.~\ref{eq:paris} holds) is fairly regular and the mean distance between striations has a well-defined characteristic value corresponding to the macroscopic crack growth rate -- indicating a one-to-one correspondence between striations and loading cycles, which has been clearly shown in an aluminum alloy using program loading~\cite{McMillan1967} as well as in many other studies~\cite{tanaka1984mechanics, cai2001striations, hershko2008assessment}.
This view is also corroborated by acoustic emission studies~\cite{deschanel2017acoustic} where the waveforms of acoustic events were observed to be nearly identical.\\

Similarly to the general characteristics of fatigue crack growth -- for example the Paris exponent $m$ -- the striation characteristics are also strongly influenced by many different factors. The loading conditions have been shown to have an effect, for example the stress ratio influences the height of striations~\cite{Uchida1999} as well as their distance~\cite{McMillan1967}. Similarly, small changes in microstructure~\cite{nygren2021influence} have been shown to affect the striation morphology and distances between them.\\

The basic reference model for striation generation is crack tip plastic slip. 
During the loading phase the crack is opened by normal stress which at the crack tip generates plastic slip activity along two symmetrical directions predicted by fracture mechanics.
In this phase the crack tip gets blunted and grows by material decohesion associated to dislocation flow into the tip or generated by stress concentration. 
As this occurs through plastic deformation, upon unloading the blunted crack tip is squished, but a new free surface remains ahead of the former crack with a new sharp tip. 
It is this step-by-step process of blunting and re-sharpening during each cycle that leaves on the crack path the kind of markings that we call striations~\cite{milella2012fatigue}. 
An extension of this approach was proposed by Laird~\cite{Laird1967TheIO} who has given a rather different interpretation of striation formation based on the so called plastic relaxation of the crack tip based on the hypothesis of plastic collapse of the crack tip during the unloading and closure phase that leads to tip concavity. 

Later Forsyth~\cite{forsyth1963fatigue} was able to classify two main types of striations: the ductile and the brittle one. Ductile striations lay on different individual planes corresponding to single grains that macroscopically form, all together, a plateau normal to the maximum tensile stress direction.
They are called ductile because the material ahead of the crack tip undergoes plastic deformations that produce the typical curved arrays by which they advance on the fracture surface. 
Brittle striations, instead, develop always on crystallographic planes, usually (100) planes and appear as concentric circles departing from the initiation site, quite often brittle inclusions. This gives brittle striations the typical flat appearance without any apparent (macroscopic) plastic deformation. A characteristic feature of brittle striations is the uniform, flat and annual ring -like propagation surface that does not propagate in single crystals but on crystallographic planes that are cleavage planes.\\

However there has also been recent imaging studies that have indicated intermittency~\cite{kokkoniemi2017intermittent, lomakin2021fatigue} and heterogeneity~\cite{carroll2013high} related to fatigue crack growth.
Especially, when tracking the crack tip during the experiment~\cite{kokkoniemi2017intermittent, lomakin2021fatigue}, the jumps in the crack length have been observed to have fat-tailed distributions.
If the accepted view of one striation corresponding to one loading cycle is true, one should then find -- in materials and loading conditions which exhibit these fat-tailed crack tip jump distributions -- a similar fat-tailed distribution of distances between striations.
Although the striations are thought to correspond to loading cycles one-to-one, several studies have shown striation spacings higher than the corresponding crack growth rate~\cite{shyam2010model, pant2016effect} or the striation spacing growing much slower than the crack growth rate~\cite{pant2016effect, moreira2007fatigue, lynch2017some}, at least for slower crack growth rates.
This can naturally be just an effect of a limited striation distance detection accuracy, although intermittency in terms of local crack advancement arrests has been proposed~\cite{gonzalez2018fractography}.
Alternatively, rough fracture surfaces deviating from the simple model of striation formation could produce surface patterns where the ridges do not directly correspond to crack arrest locations.
Similar problems of reconciling rough surfaces and periodic striations have also been encountered in geophysics in the context of ridges in surface topography~\cite{lovejoy2007scaling}.\\

In this paper we perform fatigue crack growth (FCG) experiments on an Al alloy and study in detail the jumps in the crack propagation. We then study the fracture surfaces of these samples with a focus on the distances between striation lines, and how these correlate with the crack growth rate. Two methods of extracting the distances between striations are presented -- one which focuses on the areas of the fracture surface exhibiting clear striation markings, and one which takes the whole imaged fracture surface into account.

\section{Material and methods}

\subsection{Microstructure}

\begin{figure}[t!]
    \centering
    \includegraphics[width=\columnwidth]{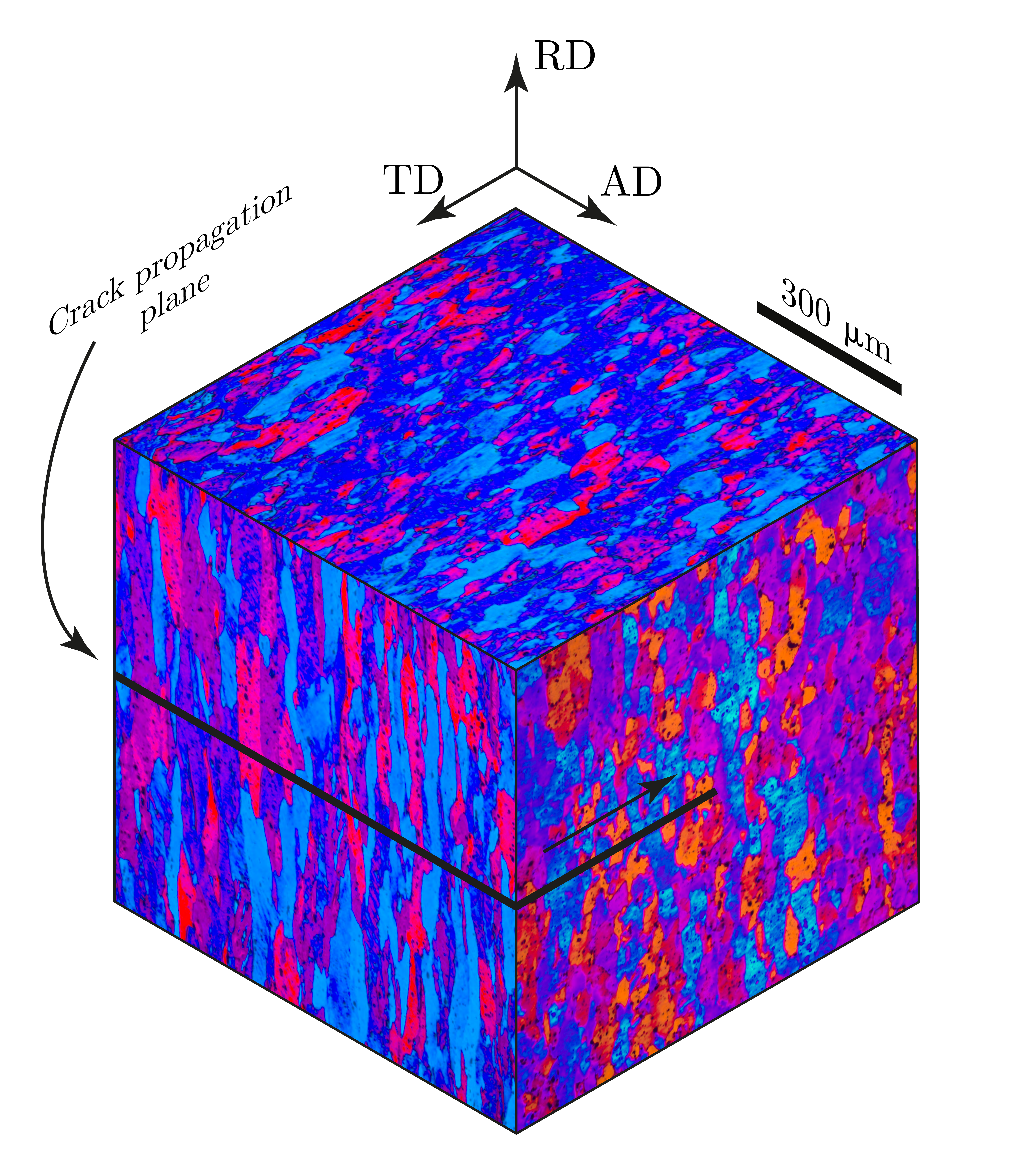}
    \caption{The grain structure of the 5005 alloy used in this study, showing elongated grains in the rolling and transverse directions. The plane of crack propagation is also indicated.}
    \label{fig:grainStructure}
\end{figure}

The material used in this study was 5005 aluminium alloy, which was provided by Alumeco Ltd. as a sheet of thickness 5~mm in a strain-hardened and partially annealed state corresponding to H24 temper. Samples for microstructural studies were cut by electrical discharge machining (EDM) and polished using a Struers Tegramin polishing machine with a final 0.3~$\mu$m OP-S suspension.
To reveal the grain structure of the material Barker' anodizing in 1.8~\% fluoboric acid water solution with 0.25~A/cm$^{2}$
and 30 V DC applied for 30~s was used. 
Further optical microscopy in polarized light was conducted using a Nikon Epiphot Inverted Metallurgical Microscope. It showed (Fig.~\ref{fig:grainStructure}) an elongated grain structure in the rolling~(RD) and transverse directions~(TD). One can see that grain size is about 300, 100 and 50~$\mu$m in rolling, transverse and axial directions (AD) respectively.

\subsection{Fatigue testing}

\begin{figure}
    \centering
    \includegraphics[width=\columnwidth]{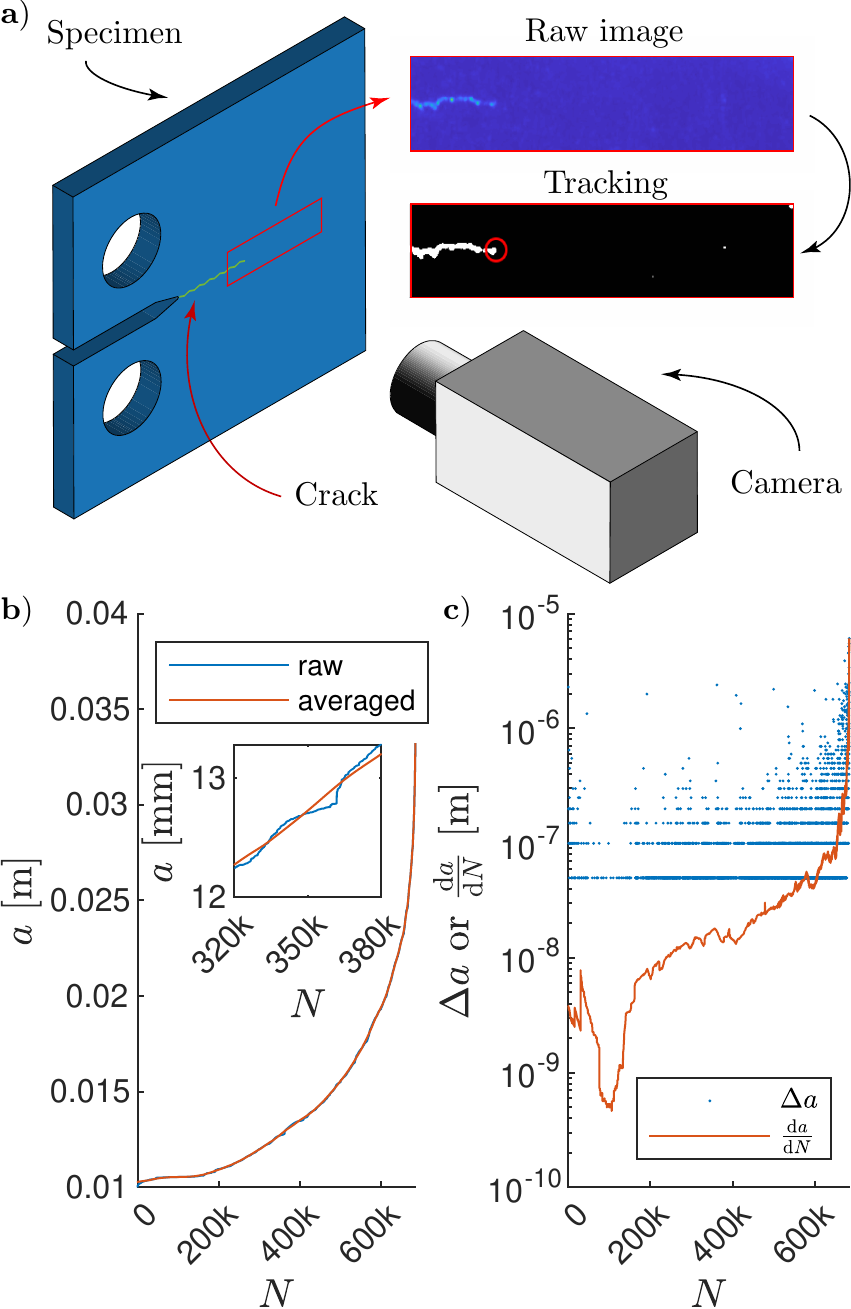}
    \caption{\textbf{a}) A representation of the imaging setup used in the fatigue tests, showing the position of the camera in relation to the sample, and a schematic of the crack tip tracking procedure.
    \textbf{b}) The resulting raw crack tip positions as a function of the number of cycles, and the result of averaging used to yield the crack tip advancement per cycle $\mathrm{d} a / \mathrm{d} N$. The inset includes a zoomed-in view, showing the jumps in the crack position.
    \textbf{c}) The jumps in the crack tip position (blue) and the crack tip advancement per cycle (orange) as a function of the number of cycles.}
    \label{fig:fcgTechnique}
\end{figure}

The FCG measurements were performed using standard compact tension (CT) specimens with a thickness of 5~mm and $W$ equal to 50~mm, cut with final notch shaping using EDM.
The orientation of the sample is such that the crack propagates in the transverse direction and the loading occurs in the rolling direction, as showed in Fig.~\ref{fig:grainStructure}.
The tests were performed using a MTS 858 hydraulic fatigue testing machine with the loading waveform being sinusoidal with a frequency of 10~Hz. 
Four different loading conditions were used and three experiments corresponding to each condition were made. First the effect of the stress ratio $R$ was studied using a constant force amplitude with a maximum force $F_{max}=1500$~N and $R$ values 0.1, 0.3, and 0.5. The effect of the force amplitude was studied using the fourth loading condition: $R=0.1$ and $F_{max} = 1300$~N.
The tests were performed as described by the ASTM~E647 standard~\cite{astmStandard} and the SIF $K$ and SIF range $\Delta K$ determined accordingly.
The yield stress of the material is 155~MPa so the 
standards requirement for predominantly elastic behavior is fulfilled up to crack lengths of $a = 0.6 W = 3.0$~cm.\\

\begin{figure*}[th!]
    \centering
    \includegraphics[width=\textwidth]{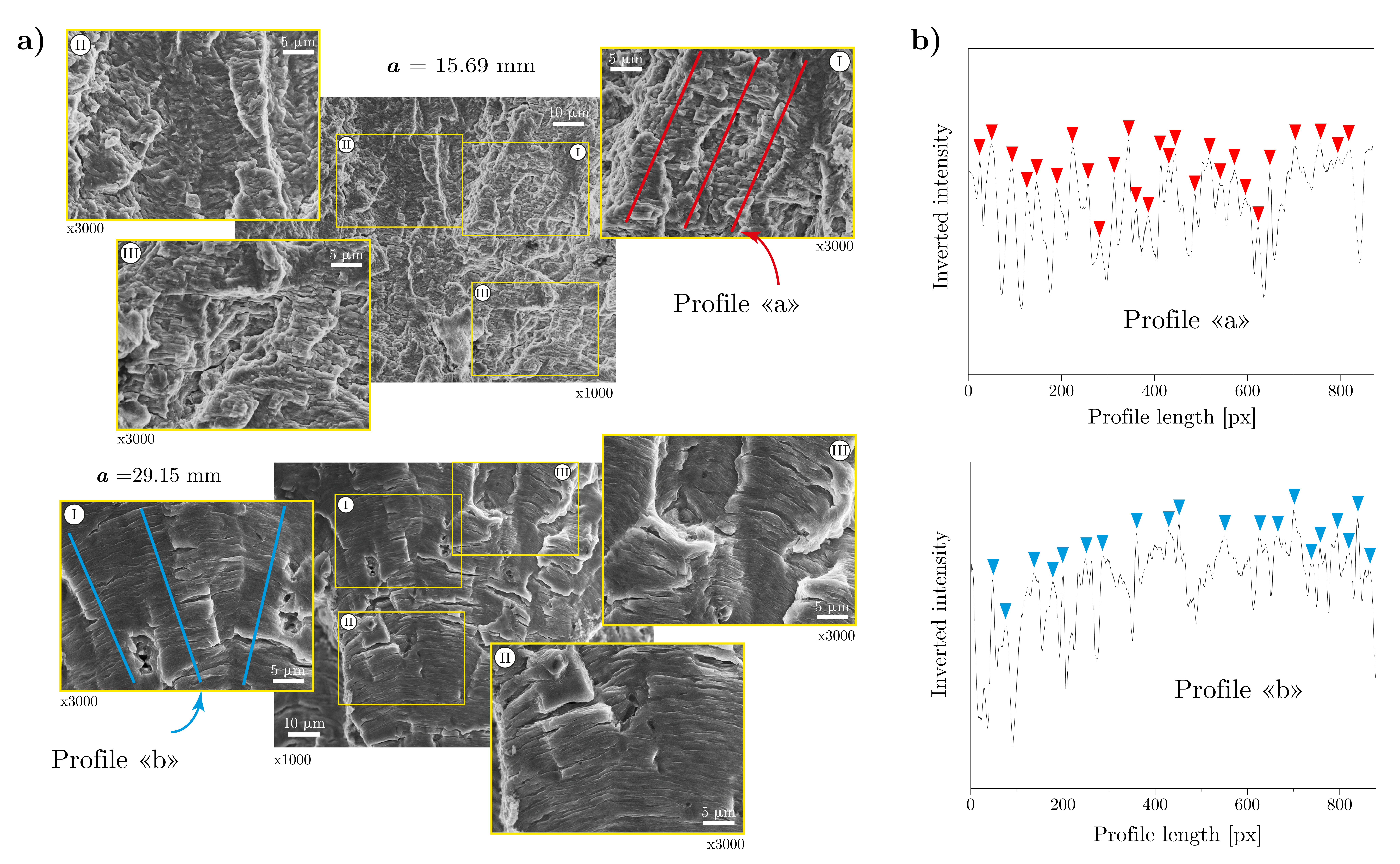}
    \caption{\textbf{a}) SEM images obtained at $a = 15.69$~mm and $a = 29.15$~mm showing the primary image and the three secondary images. The lines on the images correspond to the lines along which the striations are tracked in the manual extraction method. 
    \textbf{b}) The profiles {\guillemetleft}a{\guillemetright} and {\guillemetleft}b{\guillemetright} corresponding to the respective lines in panel a, and the results of the peak finding (triangles) used in the manual extraction method to extract the striation lines.}
    \label{fig:manualExtraction}
\end{figure*}

The crack length measurements were performed optically using the experimental setup shown in Fig.~\ref{fig:fcgTechnique}a.
The sample was imaged from the side using a Canon~EOS~R digital camera with an imaging frequency of 0.25~Hz, corresponding to 40~loading cycles between images.
The size of the obtained images was 6720~px $\times$ 4480~px, corresponding to a resolution of 4.5~$\mu$m per pixel.
Before the experiments, the sample surface was polished to a mirror-like condition using a 1~$\mu$m diamond Struers DP-Suspension as the final step. To illuminate the crack a ring LED lamp creating an analog of dark field imaging was mounted on Canon MP-E 65mm f/2.8 macro lens ($\times$1-$\times$5 magnification) attached to the camera. 

To track the advancement of the crack tip an image processing algorithm (used also in Ref.~\cite{lomakin2021fatigue}) was developed and implemented in MATLAB software. The crack tip is defined as the edge pixel of the contrast area in the crack tip region defined based on the previous image. The contrast image was obtained as an outcome of a binarization procedure using local background intensity as the threshold value. Crack tip position was tracked relative to a stationary surface defect to exclude measurement distortions brought on by the specimen shift. Sequential processing of the images allows one to track the crack length $a$ as a function of cycle number $N$ (resulting curve shown in Fig.~\ref{fig:fcgTechnique}b).

The accuracy of this tracking method is of the order of one pixel, and this small noise is filtered by requiring that the crack length increases monotonously. This is done by constructing a monotonic upper and lower envelope for the signal and taking their average to correspond to the crack length $a$. For determining the crack tip advancement per cycle $\mathrm{d} a / \mathrm{d} N$ a moving average of the raw crack length values is taken, to smoothen the signal, after which numerical differentiation is applied (see Fig.~\ref{fig:fcgTechnique}b for comparison between the raw and averaged curves).
From the raw crack tip position $a$ we also extract the crack tip jumps $\Delta a$, which are just the difference in the crack tip position between consecutive images divided by the number of cycles between the images, which here is 40. This normalization is done just to make the jump size values comparable to the $\mathrm{d} a / \mathrm{d} N$ and does not reflect an actual improvement in the tracking resolution. 
In simple terms, $\Delta a$ denotes the jumps seen in the crack tip and $\mathrm{d} a / \mathrm{d} N$ denotes the same signal, but is sufficiently smoothened to enable the plotting of the Paris curves. This is illustrated in Fig.~\ref{fig:fcgTechnique}c where both of these are plotted. In the beginning of the test the values of $\Delta a$ are much higher, as during many cycles no crack advancement is observed.
The fitting of the jump size distributions, as well as later the striation spacing distributions, is done using maximum likelihood estimation~\cite{baro2012analysis}.

\subsection{Fractography}

Three fatigue samples were picked for the fractography studies, two with $F_{max} = 1500$~N ($R=0.1$ and $R=0.5$) and one with $F_{max}=1300$~N ($R=0.1$). We performed SEM imaging of the \emph{post-mortem} fracture surfaces of the specimens. The SEM images were recorded using JEOL~JSM-7500FA microscope operated at 15~kV, and with an Everhart-Thornley secondary electron detector to reveal a surface roughness contrast.
To study the evolution of the striation line structures as a function of the crack length $a$, the following procedure was followed: primary images were taken at intervals of around 500~$\mu$m with $\times$1k magnification, and after a visual examination of the images, three regions of interest containing striation lines were recorded with a higher magnification of $\times$3k. This procedure is illustrated in Fig.~\ref{fig:manualExtraction}a.\\

\subsubsection{Roughness}

\begin{figure}[t!]
    \centering
    \includegraphics[width=\columnwidth]{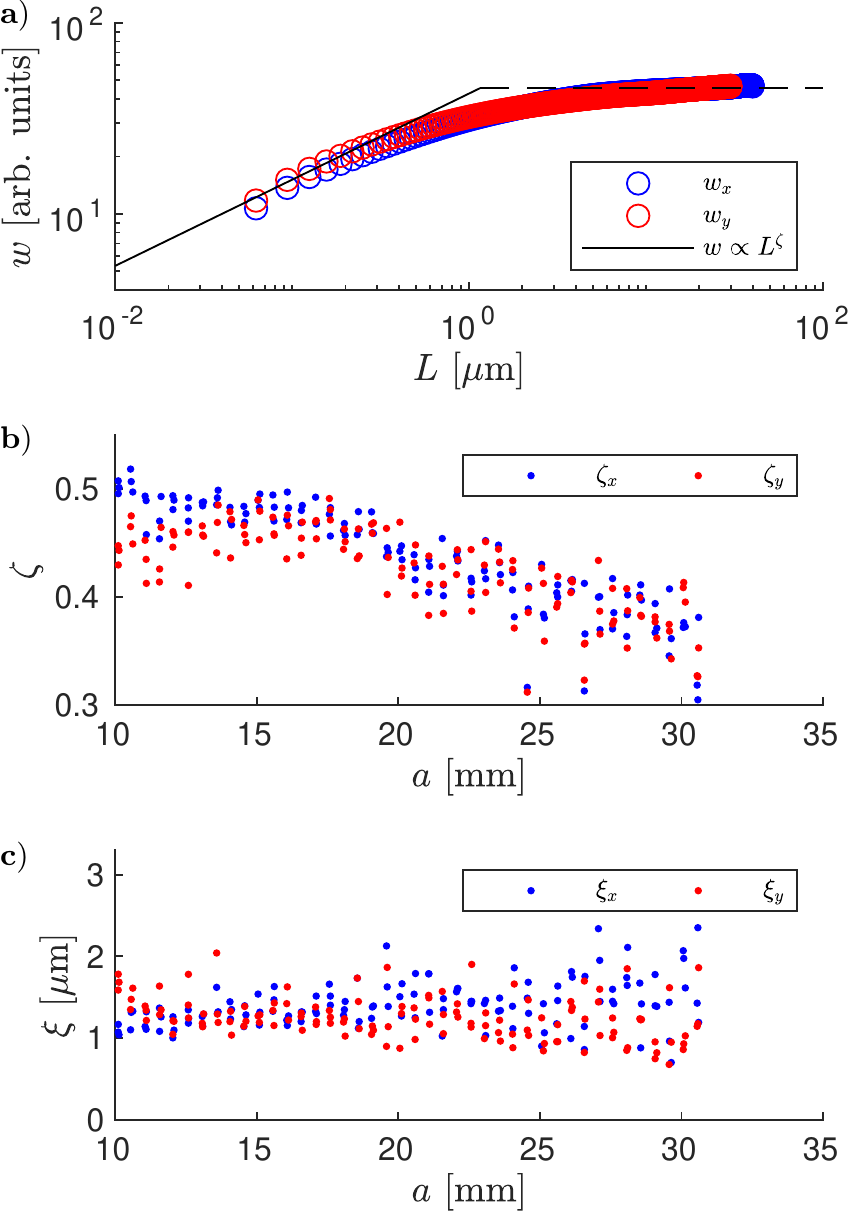}
    \caption{\textbf{a}) An example of the surface roughness extracted from one image in the two perpendicular directions $w_y$ and $w_y$ as a function of the length scale $L$, when the image is rotated so that the crack propagation occurs in the $y$-direction. The solid black line corresponds to the fitted power-law (for $w_y$) and the dashed line to the saturation value.
    \textbf{b}) The power-law exponent $\zeta$ extracted from the roughness curves as a function of the crack length $a$.
    \textbf{c}) The correlation length $\xi$ extracted from the roughness curves as a function of the crack length $a$.
    }
    \label{fig:roughness}
\end{figure}

We compute a roughness measure of the SEM images at length scale $L$ as the root-mean-square deviation of the image intensity
\begin{equation} \label{eq:roughness}
    w(L) = \sqrt{\left\langle \left( I - \left\langle I \right\rangle_L \right)^2 \right\rangle_L}
\end{equation}
where $\langle \cdot \rangle_L$ denotes an average over the length scale $L$.
We do this in both the spatial directions to yield $w_x$ and $w_y$, which are shown in Fig.~\ref{fig:roughness}.

It is important to note that SEM images do not correspond to a height map of the fracture surface and thus the roughness measure of Eq.~\ref{eq:roughness} is not an actual surface roughness. The values measured as SEM image intensity actually look more like the local slope or height variation of the fracture surface \cite{vernede2015turbulent}. However the roughness measure still holds valuable information about the structure of the fracture surface.\\

As the roughness for small length scales seems to scale as a power-law, we fit a relation $w \propto L^\zeta$ to the data. Additionally we see to which value the roughness saturates to on large length scales (by taking the average of $w$ over the latter half of our available length scales) and compute the correlation length $\xi$ by seeing at which length scale the fitted power-law would achieve this value. The process is visually illustrated in Fig.~\ref{fig:roughness}a.\\

As the crack propagation direction (the direction perpendicular to the striation lines) in the SEM images doesn't exactly coincide with the $y$-direction -- e.g. the lines drawn in Fig.~\ref{fig:manualExtraction}a do not align with the $y$-direction -- these directions are somewhat arbitrary. We can however utilize the roughness information to mitigate this problem, by rotating the image by an angle~$\theta$ and recomputing $w_x$ and $w_y$. 
We notice that the angle~$\theta$ corresponding to the angle where the crack propagation (on average) aligns with the $y$-direction seems to be the one that minimizes the ratio $w_y/w_x$.
This angle is computed for each length scale $L$ and we take the average over the intermediate length scales (from 20 to 200 pixels, corresponding to from 0.6~$\mu$m to 6.3~$\mu$m).
This observation and the aligned images are utilized in the automatic striation extraction method explained later.

\subsubsection{Manual striation extraction}

The first method for striation extraction is the manual method presented here (similar to the one used in Ref.~\cite{shyam2010model}). As the crack advancement direction does not always coincide with the $y$-direction of the images, and there are other features present in the images in addition to the striation lines, we have manually marked the parts in the images containing striations. In practice this means drawing lines on the images (shown in Fig.~\ref{fig:manualExtraction}a) and performing a peak-finding procedure (shown in Fig.~\ref{fig:manualExtraction}b) to yield the positions of the striation lines, which lay perpendicular to the drawn lines. The peak finding is done for the inverted intensity, as we are tracking the depressions on the sample surface.
The distance between striations~$\ell$ is then directly computed from the distances between the peaks, yielding a set of $\ell$~values corresponding to each image (which in turn correspond to a crack length~$a$).
In total, 35865 striation line measurements along fracture surfaces of the specimens were made.

\begin{figure}[t!]
    \centering
    \includegraphics[width=\columnwidth]{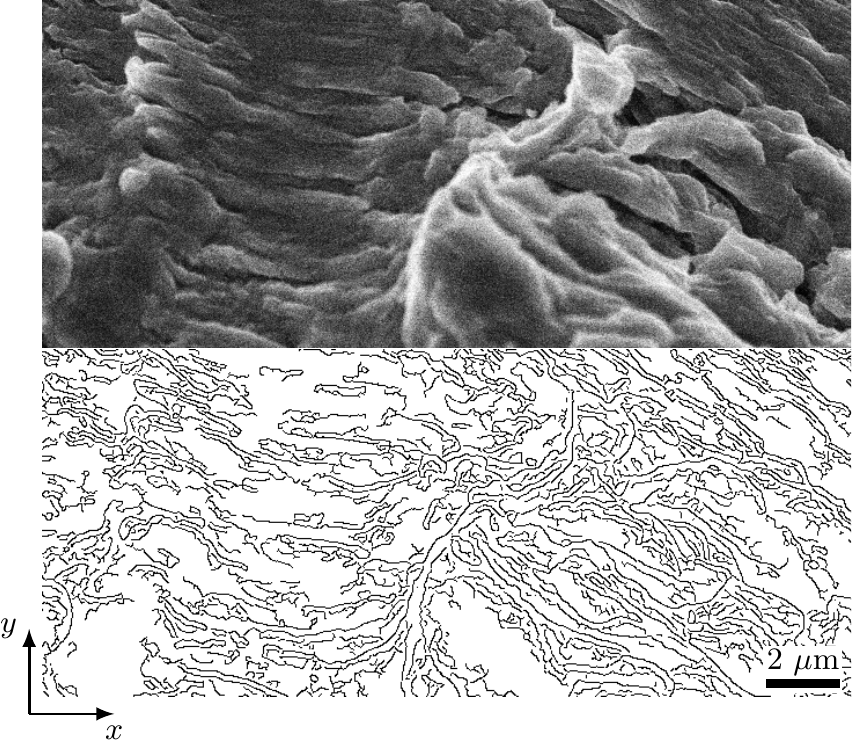}
    \caption{An example of the automatic striation extraction. The top image is the raw (rotated) SEM image and the bottom one the result after striation extraction. The distance between striations~$\ell$ is then the distance between lines in the bottom image in the $y$-direction, computed for each pixel column of the image.
    }
    \label{fig:autoExtraction}
\end{figure}

\subsubsection{Automatic striation extraction}

As the goal of the present study is to see if the distances between striation lines follow the same fat-tailed distributions as observed for the jumps in the crack tip, the manual tracking method presented might be problematic.
It is feasible that in selecting the parts of the image to be considered, the human tendency is to select the parts where the striation line spacing is fairly regular, yielding a mismatch between the crack tip jump results and the striation spacing.
This type of issue related to selection of areas and length scales has in other contexts been dubbed the ''phenomenological fallacy''~\cite{lovejoy2007scaling}.\\

To try to get rid of this possible issue, we have also implemented an automatic striation extraction method. It involves first rotating the image so that the crack propagation direction (on average) aligns with the $y$-direction in the images. Then a simple Canny edge detector~\cite{canny1986computational} algorithm is employed (again on the inverted image intensity, as we are trying to track the depressions) to yield the striation lines shown in Fig.~\ref{fig:autoExtraction}. We then go through each column of pixels in the image and extract the striation distances as the distance between these lines generated by the edge detection.
This method of striation distance determination has some obvious issues with multiple counting of the same distances, but this choice was made because the structures resulting from edge detection have a much more complicated structure than just straight lines.

\section{Results}

\subsection{Fatigue testing}

\begin{figure}[t!]
    \centering
    \includegraphics[width=\columnwidth]{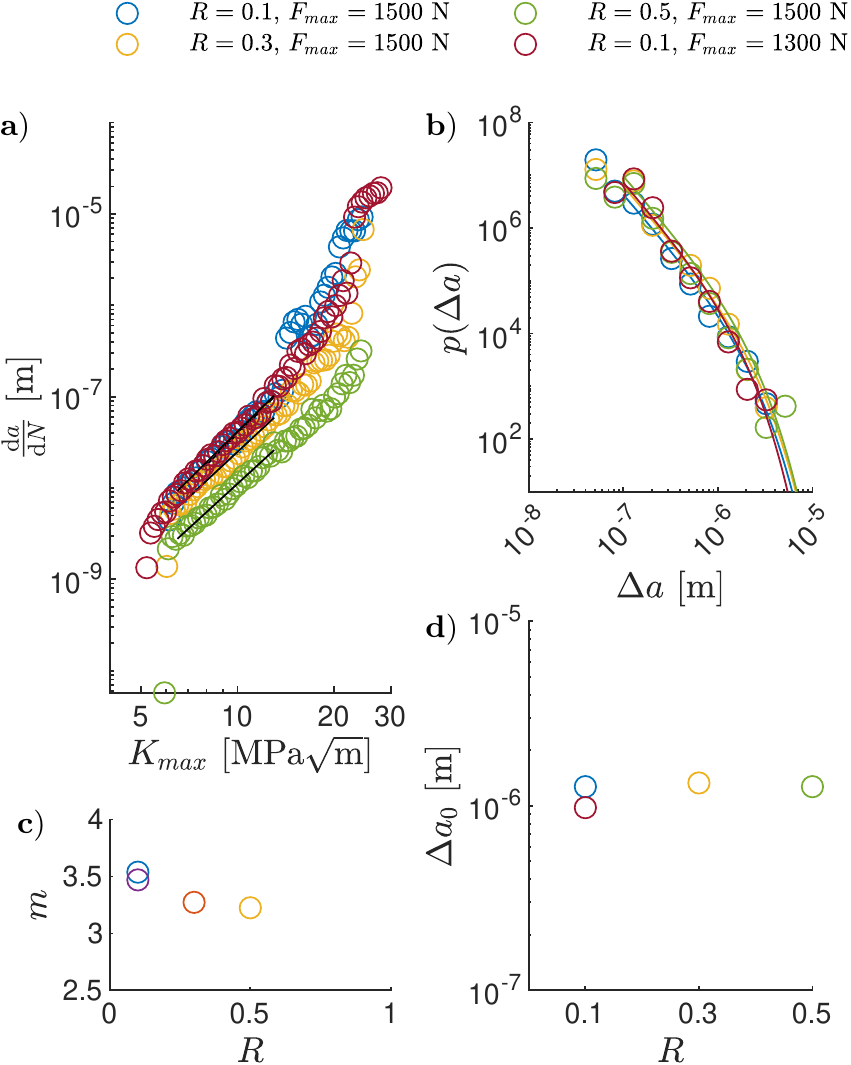}
    \caption{\textbf{a}) The Paris curves for each of the loading conditions and the exponent fits (black lines) according to Eq.~\ref{eq:paris} to each of them in the Paris regime.
    \textbf{b}) The probability distributions of the crack advancement jump sizes $\Delta a$ for each of the loading conditions. The lines (with corresponding colors) are maximum likelihood fits to Eq.~\ref{eq:distribution} with a fixed exponent $\gamma = 2$.
    \textbf{c}) The fitted values of the exponent $m$ (Eq.~\ref{eq:paris}) as a function of the stress ratio $R$.
    \textbf{d}) The fitted cutoff sizes of the crack advancement jumps $\Delta a_0$ (Eq.~\ref{eq:distribution}) as a function of the stress ratio $R$.
    }
    \label{fig:fcgResults}
\end{figure}

The fatigue testing results align with previous results reported in Ref.~\cite{lomakin2021fatigue} for aluminum 1050 alloy. In the Paris plots (Fig.~\ref{fig:fcgResults}a) we see the Paris--Erdogan law holding for each of the loading conditions in the region from $K_{max}=6.5$~MPa$\sqrt{\mathrm{m}}$ to $K_{max}=13$~MPa$\sqrt{\mathrm{m}}$ with exponent values slightly over three. For SIF values less than this we see the typical threshold regime behavior of rapidly decreasing crack growth rate, and above this the crack growth rate increases quickly as the critical SIF is approached.

\begin{figure*}[th!]
    \centering
    \includegraphics[width=\textwidth]{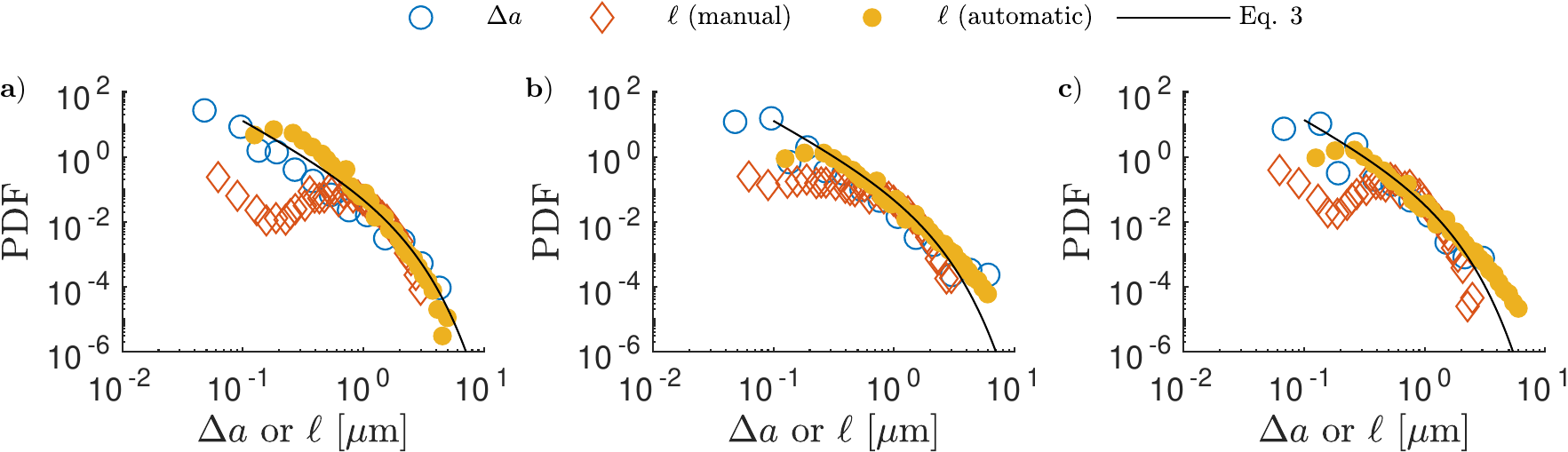}
    \caption{The distributions of three quantities: crack advancement jump size $\Delta a$ and the distance between striations~$\ell$ measured with manual and automatic tracking. The three plots correspond to different loading conditions, \textbf{a}) $F = 1500$~N and $R=0.1$, \textbf{b}) $F = 1500$~N and $R=0.5$, and \textbf{c}) $F = 1300$~N and $R=0.1$.}
    \label{fig:striationDist}
\end{figure*}

The almost linear change in the exponent~$m$ as a function of the stress ratio~$R$ (see inset of Fig.~\ref{fig:fcgResults}a) is much smaller than in Al 1050 alloy and the exponent does not seem to tend to zero at the creep limit $R \to 1$. The change in the maximum force from 1500~N to 1300~N has a negligible effect on the exponent $m$.\\

When looking at the crack tip jump size distributions~$p(\Delta a)$ (Fig.~\ref{fig:fcgResults}b) we see fat-tailed distributions spanning around two orders of magnitude. They can be interpreted, similarly as in the Al 1050 alloy~\cite{lomakin2021fatigue}, as power-laws with an exponential cutoff
\begin{equation} \label{eq:distribution}
    p(\Delta a) \propto \Delta a^{-\gamma} \exp\left( - \frac{\Delta a}{\Delta a_0} \right)
\end{equation}
where $\gamma$ is an exponent around 2 and $\Delta a_0$ denotes the cutoff scale. The solid lines in Fig.~\ref{fig:fcgResults}b show maximum likelihood fits of the cutoff scales~$\Delta a_0$ (for a fixed $\gamma = 2$) for each of the loading conditions. One can clearly see (Fig.~\ref{fig:fcgResults}c) that the differences in the jump size cutoff are very small.

\subsection{Roughness}

When the images are rotated so that the crack propagation occurs (on average) in the $y$-direction, using the aforementioned method, we observe (see Fig.~\ref{fig:roughness}a) for small distances a power-law scaling $w \propto L^\zeta$ where the exponent $\zeta$ is initially around 0.5 and decreases to around 0.4 with increasing $a$ (see Fig.~\ref{fig:roughness}b). 
This is significantly lower than the values of the actual roughness exponent observed for cracked aluminum alloys \cite{hinojosa2000roughness, hinojosa2002self, barak2019correlating}. There is very little difference between the exponent values in the two directions $\zeta_x$ and $\zeta_y$, although initially $\zeta_x$ is slightly larger.
This power-law scaling is observed up to a correlation length $\xi$, which is approximately constant for the whole experiment (see Fig.~\ref{fig:roughness}c).

\subsection{Striations}

The striations we observe (see Fig.~\ref{fig:manualExtraction}a) are typical ductile striations~\cite{milella2012fatigue} residing in channel-like areas separated by ridges. By comparing the striation morphology at the beginning of the experiment (the images corresponding to $a=14.69$~mm in Fig.~\ref{fig:manualExtraction}a) and in the end of the experiment (the images corresponding to $a=29.15$~mm in Fig.~\ref{fig:manualExtraction}a) one sees that the channels seem to get wider as the experiment progresses, as seen previously in aluminum alloys~\cite{zhang2019microstructure}.
However, when comparing the automatically extracted striation spacings $\ell$ in the $x$ and $y$-directions we see just a linear dependence.\\

The striation spacings extracted manually do not follow a fat-tailed distribution, as can be seen in Fig.~\ref{fig:striationDist}. Instead they follow fairly closely an exponential distribution of the form $p(\ell) \propto \exp(-\ell/\ell_0)$, where the scale parameter $\ell_0$ aligns well with the cutoff scale $\Delta a_0$ of the crack tip jump size distributions, i.e. the manual extraction method manages to capture striations comparable in size to the exponential tail of the optically measured crack tip jumps -- the largest jumps.

When the automatic extraction is performed, one indeed sees a fat-tailed distribution of striation spacings (see Fig.~\ref{fig:striationDist}), spanning around two orders of magnitude, as do the crack tip jump sizes. We have done a maximum likelihood fit to Eq.~\ref{eq:distribution} with a fixed $\gamma=2$ and this distribution fits the data reasonably well. Similarly to the crack tip jump sizes and the manually extracted striation spacings, the effect of loading conditions on the the cutoff scale is very small.
In the very end of the tail of the distribution (sizes of several micrometers) the discrepancies between the distributions are larger. This is due to the statistics -- in these bins there are very few datapoints.\\

\begin{figure}[th!]
    \centering
    \includegraphics[width=\columnwidth]{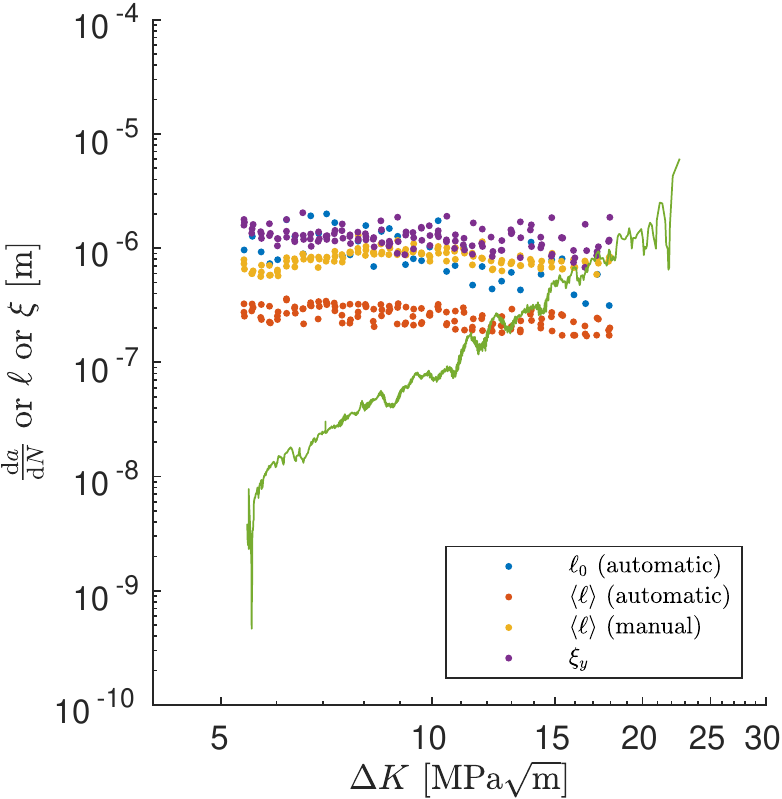}
    \caption{The Paris curve of one experiment ($R=0.1$, $F_{max} = 1500$~N) with the the evolution of the microscopic quantities as a function of the SIF range superimposed. The quantities are the average striation spacing $\langle \ell \rangle$ determined by the manual and automatic extraction methods, the cutoff scale fitted to the automatically extracted striation spacings $\ell_0$ and the correlation length $\xi_y$ in the crack propagation direction determined by the roughness analysis.}
    \label{fig:parisAndMicro}
\end{figure}

When looking at the evolution of the microscopic observables -- the cutoff scale~$\ell_0$ from the automatically extracted striation spacings, the mean striation spacing~$\langle \ell \rangle$ extracted using both the manual and automatic methods, and the correlation length in the crack propagation direction $\xi_y$ -- one can clearly see (Fig.~\ref{fig:parisAndMicro}) that they stay approximately constant for the whole duration of the experiment. During the same experiment, the macroscopic crack growth rate varies around four orders of magnitude. 
If the one-to-one correlation between striation spacings and the macroscopic crack growth rate held, one would expect the curves to have the same slope in this log-log-plot.
Three of these microscopic observables ($\ell_0$ from automatic extraction, $\langle \ell \rangle$ from manual extraction, and $\xi_y$) have roughly the same value -- corresponding roughly to the macroscopic crack growth rate at the end of the experiment -- and the $\langle \ell \rangle$ from the automatic extraction has a slightly smaller value due to the automatically extracted striation spacings spanning a larger range of values.

\section{Conclusions}
Starting with crack propagation, comparing the FCG results with similar ones for Al 1050 alloy~\cite{lomakin2021fatigue}, we find that the change in the Paris law exponent is much smaller with increasing $R$. We suggest that this difference can be attributed to a decrease in the sample plasticity, as Al 5005 is as an alloy significantly more brittle than Al 1050. Additionally we note that a change in the Paris exponent makes the usual models~\cite{kujawski2001fatigue,noroozi2005two,huang2008engineering} for the effect of the stress ratio $R$ not effective here, as they only apply a $R$ dependent prefactor to the SIF value, therefore not changing the slope of the curves whereas here the slopes change.

The statistical analysis of the crack tip advancement on a microscopic scale (down to a few microns) shows a fat-tailed distribution of crack tip jump sizes. The distribution can be modelled as a power-law with an exponential cutoff where the exponent is around~2 for all the loading conditions. The changes in the cutoff scale between different loading conditions are too small to make meaningful distinctions but this seems logical with the changes in the Paris curves also being small. This exponent~2 also agrees with the Al-1050 results, hinting at a degree of universality, but much more statistics would be needed for definitive conclusions. We can not for example exclude some other fat-tailed distributions, such as the the streched exponential. 
However, a plausible explanation for this exponent can be found from previous analysis of aluminum fracture surfaces~\cite{vernede2015turbulent} where the distribution of microcrack sizes was found the be power-law distributed with an exponent around~2. If one assumes the coalescence of these microcracks as the primary crack growth mechanism~\cite{paun2003morphology}, this would then lead to a crack tip jump distribution with the same exponent.\\

On the fractography side we study the roughness extracted from the images, which on small scales scales as a power-law of the length scale with an exponent slightly below 0.5. This exponent value also decreases during the experiment.
When interpreting these results, one should however take into account that we are not measuring the surface height, only the SEM image intensity.

We then introduce two methods of striation spacing extraction: the manual and the automatic one.
The manual extraction method focuses on the features that can clearly be identified as striations and the automatic one considers all depressions on the fracture surface.
The two extraction methods agree roughly statistically on the striation spacings when the spacing is above 1~$\mu$m.
The manual one neglects the smaller distances between striations, that seem to be at least statistically captured by the automatic method. The automatic method also produces a striation spacing distribution which roughly matches the one seen for the crack tip jump sizes.

However none of the microscopic quantities measured (the striation spacings or the correlation length associated with roughness) capture features that would correspond to the macroscopic crack growth rate spanning multiple orders of magnitude. The only microscopic quantity where any significant change during the experiment can be seen is the small length scale power-law exponent $\zeta$ of the roughness.
If one accepts that the spacing between striation lines does not evolve during the experiment, one could envisage a mechanism of striation channel widening to account for the changes in the crack growth rate. The motivation for this approach comes from the clearly different striation morphology when comparing the beginning and end of the experiment.
However, the results of our automatic tracking do not support this mechanism but the limited field of view of our SEM imaging also does not allow for a full characterization of this phenomenology. One should also note that this type of behavior would deviate from the constant shape or statistics of avalanche fronts observed for other crack avalanching systems in non-fatigue loading~\cite{maaloy2006local,bonamy2008crackling,laurson2010avalanches}.\\

We have shown that the striation spacings in intermittent crack growth have fat-tailed distributions, in accordance with the one-to-one correspondence between striations and loading cycles. The manual extraction method only captures the striations with a well-defined mean value -- corresponding to the largest jumps -- but the automatic one also shows a plethora of features with narrower spacings. These statistically match the crack tip jump sizes, but it is a matter of terminology if these features should actually be called striations or apparent striations~\cite{gonzalez2018fractography, devries2010counting}.
One should note that the images captured represent only a part of the fracture surface and clear striation-like markings are seen only on parts of the images.
Generally crack propagation seems to be a much more complex phenomenon than just crack advancement lines with a well-defined spacing, as illustrated by the inconsistency between the striations and the macroscopic crack growth rate.
On this rough and complicated fracture surface, the observed ridges might not correspond to crack arrest locations implied by the simplified model for striation formation.\\

Further work should be done to see if the observed universal features of intermittent crack growth in fatigue extend to a wider variety of materials and to explore the effects with much better statistics. 
It would be interesting to see if the area fraction of the fracture surface exhibiting striation-like markings is correlated with the intermittency of crack propagation.
One should note that in Al alloys increasing the Mg content further is known to introduce dynamic strain aging, which might significantly complicate the crack propagation dynamics.
The validity of the striation line extraction methods introduced here should also be verified with other materials. By performing direct measurements of the fracture surface topography the possible connection to the microcrack size distribution should be explored.

\begin{acknowledgements}
M.J.A. and T.M. acknowledge support from the European Union Horizon 2020 research and innovation programme under grant agreement No 857470 and from European Regional Development Fund via Foundation for Polish Science International Research Agenda PLUS programme grant No MAB PLUS/2018/8.
M.J.A. acknowledges support from the Academy of Finland (Center of Excellence program, 278367 and 317464).
J.K. acknowledges the funding from Academy of Finland (308235) and Business Finland (211715).
I.V.L. acknowledges the funding from Academy of Finland (341440 and 346603).
The authors acknowledge the computational resources provided by the Aalto University School of Science ``Science-IT'' project and the provision of facilities and technical support by Aalto University at OtaNano -- Nanomicroscopy Center (Aalto-NMC).
\end{acknowledgements}

%

\end{document}